\RequirePackage{fix-cm}
\documentclass[smallextended]{svjour3}       
\smartqed  
\usepackage{graphicx}
\usepackage{amsfonts}
\begin{document}

\title{Quantum noise of electron-phonon heat current}


\author{ Jukka P. Pekola        \and  Bayan Karimi
}


\institute{Jukka P. Pekola \at
              Low Temperature Laboratory, Department of Applied Physics, Aalto University School of Science, P.O. Box 13500, 00076 Aalto, Finland \\
              \email{jukka.pekola@aalto.fi}           
           \and
           Bayan Karimi \at
              Low Temperature Laboratory, Department of Applied Physics, Aalto University School of Science, P.O. Box 13500, 00076 Aalto, Finland \\
              \email{bayan.karimi@aalto.fi}
}
\date{Received: date / Accepted: date}

\maketitle

\begin{abstract}
We analyze heat current fluctuations between electrons and phonons in a metal. In equilibrium we recover the standard result consistent with the fluctuation-dissipation theorem. Here we show that heat current noise at finite frequencies, remains non-vanishing down to zero temperature. We briefly discuss the impact of electron-phonon heat current fluctuations on calorimetry, in particular in the regime of single microwave-photon detection.
\end{abstract}

\section{Introduction}
\label{intro}
At low temperatures, phonons provide the heat bath to which electrons couple weakly in a mesoscopic electron circuit. Due to fast electron-electron relaxation \cite{pothier}, electrons typically obey Fermi-Dirac distribution even under non-equilibrium conditions with a well-defined temperature $T_e$ that can differ from the temperature $T_p$ of the phonons. For clean metals in three-dimensional structures, the relaxation rate of electrons to the temperature of phonons scales as $T_e^{-3}$ \cite{gantmakher,roukes,wellstood}. This result is based on the electron-phonon coupling arising from standard deformation potential model \cite{fetter}. For an arbitrary temperature difference (as long as the model is valid) the average heat current between electrons and phonons scales as $T_e^5 -T_p^5$ in the same model, and this result is widely observed in experiments at sub-kelvin temperatures.
\begin{figure}[!h]
\centering
 \includegraphics[width=0.4\textwidth]{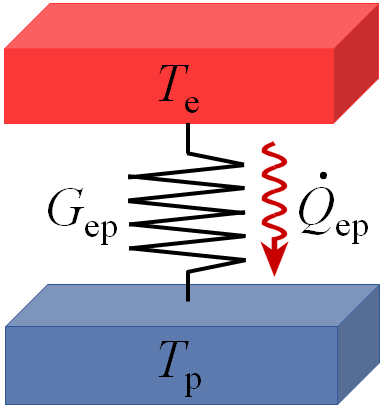}
\caption{Normal metal with temperature $T_e$ coupled to phonon bath ($T_p$) via electron-phonon coupling with thermal conductance $G_{ep}$.}
\label{fig:1}       
\end{figure}
Detection of single quanta of radiation by calorimetric means has become popular over the past years \cite{Eisaman,JP}. In this context, not only the average heat current but also its fluctuations are important. They set the fundamental bound of minimum detectable energy of the radiation quantum. Here we re-visit the standard results of heat current fluctuations under equilibrium and non-equilibrium conditions, which influence, e.g., single-photon detection in the microwave regime \cite{Dima,golwala}. As the main result we present electron-phonon heat current noise at finite frequencies, and realize that it is non-vanishing even in the zero temperature limit. This result has interesting implications in terms of the fluctuation-dissipation theorem for heat in the quantum regime, the topic discussed over the past few years in the context of electron transmission in tunnel contacts and through general scatterers \cite{pekolaPRL}. 

\section{Description of the system and heat current operators}
\label{sec:2}
A sketch of the system is schematically shown in Fig. \ref{fig:1}. The normal metal is thermally coupled to the local phonon bath with thermal conductance $G_{ep}$ at temperature $T_{\rm p}$. The total Hamiltonian describing the system and the environment is given by
\begin{equation} \label{hamiltonian}
H=H_e+H_p+H_{\rm ep},
\end{equation}
where $H_e, H_p$ are the Hamiltonians of the electrons and phonons, respectively, and $H_{\rm ep}$ is the coupling between them. The unperturbed Hamiltonian $H_0 = H_e+H_p$ can be written as 
\begin{equation}\label{unperturbed hamiltonian}
H_0=\sum_k \epsilon_k a_k^\dagger a_k+\sum_q \hbar\omega_q c_q^\dagger c_q,
\end{equation}
where the first part describes electron states with energy $\epsilon_k$, momentum $k$, and $a_k^\dagger$ and $a_k$ are the corresponding creation and annihilation operators. With analogous notation, the second part shows the Hamiltonian of phonons with eigenenergies $\hbar \omega_q$, wavevector $q$, and bosonic creation and annihilation operators $c_q^\dagger$ and $c_q$. The coupling term as a perturbation of the system has the following form in a metal \cite{fetter}
\begin{equation} \label{perturbation}
H_{\rm ep}= \gamma \sum_{k,q} \omega_q^{1/2}(a_k^\dagger a_{k-q}c_q + a_{k-q}^\dagger a_{k}c_q^\dagger).
\end{equation}
Here, the magnitude of $\gamma$ depends on the material properties of the system. The operator of heat flux from the electron system to phonons due to ep coupling is 
\begin{equation} \label{heatp2}
\dot H_p=\frac{i}{\hbar}[H_{\rm ep},H_p]=i\gamma \sum_{k,q} \omega_q^{3/2}( a_k^\dagger a_{k-q}c_{q} - a_{k-q}^\dagger a_{k}c_q^\dagger),
\end{equation}
where we used the commutation relations for the bosonic operators, $[c_q,c_q^\dagger c_q]=c_q$ and $[c_q^\dagger,c_q^\dagger c_q]=-c_q^\dagger$.
Similarly we find the operator for heat flux to electron system 
\begin{equation} \label{heate}
\dot H_e= -\frac{i\gamma}{\hbar} \sum_{k,q} \omega_q^{1/2}(\epsilon_k - \epsilon_{k-q})(a_{k}^\dagger a_{k-q}c_{q}-a_{k-q}^\dagger a_{k}c_{q}^\dagger).
\end{equation}

\section{Fluctuations of heat current}
\label{sec:3}
In order to find the heat current fluctuations, we evaluate the correlator $\langle \mathfrak{I}(t) \mathfrak{I}(0)\rangle$, where $\mathfrak{I}\equiv \frac{1}{2}(\dot H_e-\dot H_p)$ is the symmetric heat current operator between electron and phonon baths. Using Eqs. (\ref{heatp2}) and (\ref{heate}) yields
\begin{equation} \label{hcurrent1}
\mathfrak I= -\frac{i\gamma}{2\hbar} \sum_{k,q} \omega_q^{1/2}(\hbar \omega_q+\epsilon_k - \epsilon_{k-q})(a_{k}^\dagger a_{k-q}c_{q}-a_{k-q}^\dagger a_{k}c_{q}^\dagger).
\end{equation}
Then we have
\begin{eqnarray} \label{ItI0}
\langle \mathfrak I(t)\mathfrak I(0)\rangle && = \frac{\gamma^2}{4\hbar^2} \sum_{k,q} \omega_q(\hbar\omega_q+\epsilon_k-\epsilon_{k-q})^2 \Big[ \langle  a_k^\dagger (t)a_k(0)a_{k-q}(t)a_{k-q}^\dagger(0)c_q(t)c_q^\dagger(0)\rangle \nonumber\\ && +
\langle a_{k-q}^\dagger (t)a_{k-q}(0)a_{k}(t)a_{k}^\dagger(0)c_q^\dagger(t)c_q(0)\rangle \Big].
\end{eqnarray}
We use the time dependence of the creation and annihilation operators, $a_k(t)=a_ke^{-i\epsilon_kt/\hbar}$ and $c_q(t)=c_qe^{-i\omega_qt}$, taking the expectation values of the products of the operators ($\langle a_{k}^\dagger a_k\rangle=f(\epsilon_k)$ and $\langle c_{q}^\dagger c_q\rangle=n(\omega_q)$), where $f(\epsilon)=(1+e^{\beta_e \epsilon})^{-1}$ and $n(\omega)=(e^{\beta_p \hbar \omega}-1)^{-1}$ are Fermi and Bose distribution for electrons and phonons, respectively, with related inverse temperatures $\beta_e=(k_BT_e)^{-1}$ and $\beta_p=(k_BT_p)^{-1}$. Integrating for the noise power $S_{\mathfrak I}(\omega)=\int dt \langle \mathfrak I(t)\mathfrak I(0)\rangle e^{i\omega t}$ leads to
\begin{eqnarray} \label{ItI0}
S_{\mathfrak I}(\omega) && = \frac{\gamma^2}{4\hbar^2} \sum_{k,q} \omega_q(\hbar\omega_q +\epsilon_k -\epsilon_{k-q})^2 \Big[ e^{i(\epsilon_k -\epsilon_{k-q}-\hbar \omega_q +\hbar\omega)t/\hbar} f(\epsilon_k)[1-f(\epsilon_{k-q})] \nonumber\\ && [1+n(\omega_q)]+e^{i(\epsilon_{k-q}-\epsilon_k+\hbar \omega_q+\hbar\omega)t/\hbar} f(\epsilon_{k-q})[1-f(\epsilon_k)]n(\omega_q)\Big].
\end{eqnarray}
Now, we integrate over time, and replace $\sum_q \rightarrow D(q)\int d^3q = \frac{\mathcal{V}}{(2\pi)^2}\int_0^\infty dq\,q^2\int_{-1}^{1} d(\cos \theta)$ in spherical coordinates, where $\theta$ is the angle between $k$ and $q$, and $\sum_k \rightarrow N(0) \int d\epsilon_k$. $N(0)$ $(D(q)=\frac{\mathcal{V}}{(2\pi)^3})$ denotes the density of states of electrons (phonons), here $\mathcal{V}$ is the volume of the system. Further, $\epsilon_{k} =\frac{\hbar^2 k^2}{2m}$, $\epsilon_{k \pm q}\simeq \epsilon_k \pm \frac{\hbar^2 k_F}{m}q\cos \theta$, where the last approximation is due to $k\simeq k_F$ and $q\ll k_F$, where $k_F$ is the Fermi wave vector and $m$ is the mass of electron. Moreover $\omega_q$ is replaced by $c_lq$, where $c_l$ is the speed of sound. Then Eq. (\ref{ItI0}) leads to
\begin{eqnarray} \label{sIw1}
&&S_{\mathfrak I}(\omega) = \frac{\gamma^2N(0)\mathcal V}{8\pi\hbar} \int_{-\infty}^{+\infty} d\epsilon_k \int_{0}^{+\infty} dq\,q^2\int_{-1}^{1} d(\cos \theta) \,\, \omega_q(\hbar \omega_q+\frac{\hbar^2 k_F}{m}q\cos \theta)^2\times \nonumber\\ &&\Big[ f(\epsilon_k)[1-f(\epsilon_{k}-\frac{\hbar^2 k_F}{m}q\cos \theta)][1+n(\omega_q)]\delta(\frac{\hbar^2 k_F}{m}q\cos \theta-\hbar\omega_q+\hbar\omega)\nonumber \\&& +
f(\epsilon_{k}-\frac{\hbar^2 k_F}{m}q\cos \theta)[1-f(\epsilon_{k})]n(\omega_q)\delta(-\frac{\hbar^2 k_F}{m}q\cos \theta+\hbar\omega_q+\hbar\omega) \Big].
\end{eqnarray}
Collecting the angle dependent terms and integrating over $\cos \theta$ and using notation $\epsilon\equiv \hbar \omega_q=\hbar c_lq$, we have
\begin{eqnarray} \label{sIw2}
S_{\mathfrak I}(\omega) =\frac{\Sigma \mathcal V}{96\zeta (5)k_B^5}\int_0^\infty d\epsilon \,\epsilon^2 &&\Big[(2\epsilon-\hbar\omega)^2\frac{1}{1-e^{-\beta_p \epsilon}}\,\,\frac{\epsilon-\hbar\omega}{e^{\beta_e(\epsilon-\hbar\omega)}-1}\nonumber\\&&+(2\epsilon+\hbar\omega)^2\frac{1}{e^{\beta_p \epsilon}-1}\,\,\frac{\epsilon+\hbar\omega}{1-e^{-\beta_e(\epsilon+\hbar\omega)}}\Big],
\end{eqnarray}
where  we have defined the electron-phonon coupling constant \cite{wellstood,cleland} as $\Sigma=\frac{12\gamma^2N(0)m\zeta (5)k_B^5}{\pi k_Fc_l^2\hbar^6}$ and $\zeta (z)$ denotes the Riemann zeta function.

One can calculate the average heat flux into the phonon bath $\dot Q_{ep} =\langle \dot H_p\rangle$ by applying the Kubo formula in the interaction picture to Eq. (\ref{heatp2}) as
\begin{eqnarray} \label{Qep}
&&\dot Q_{ep}=-\frac{i}{\hbar}\int_0^\infty dt'\langle [\dot H_p(t),H_{ep}(t')]\rangle.
\end{eqnarray}
We obtain then the known result
\begin{eqnarray} \label{zeroT}
&&\dot Q_{ep} = \Sigma \mathcal V (T_e^5-T_p^5).
\end{eqnarray}
For the noise spectrum at equal temperatures, $T_e=T_p$, and $\omega=0$, we obtain
\begin{equation}\label{zeroTS}
S_{\mathfrak I}(0) = 10\Sigma \mathcal V k_BT^6,
\end{equation}
which is again the well known classical result \cite{Dima}. In this regime $\dot Q_{ep}$ yields the thermal conductance between electrons and phonons, $G_{ep}$, by differentiation with respect to $T_e$ at $T=T_e=T_p$. We have $G_{ep}=d\dot Q_{ep}/dT_e=5\Sigma\mathcal{V}T^4$, which satisfies the fluctuation-dissipation relation
\begin{equation}
S_{\mathfrak I}(0)=2 k_B T^2 G_{ep}.
\end{equation}
\begin{figure}[!h]
\centering
 \includegraphics[width=\textwidth]{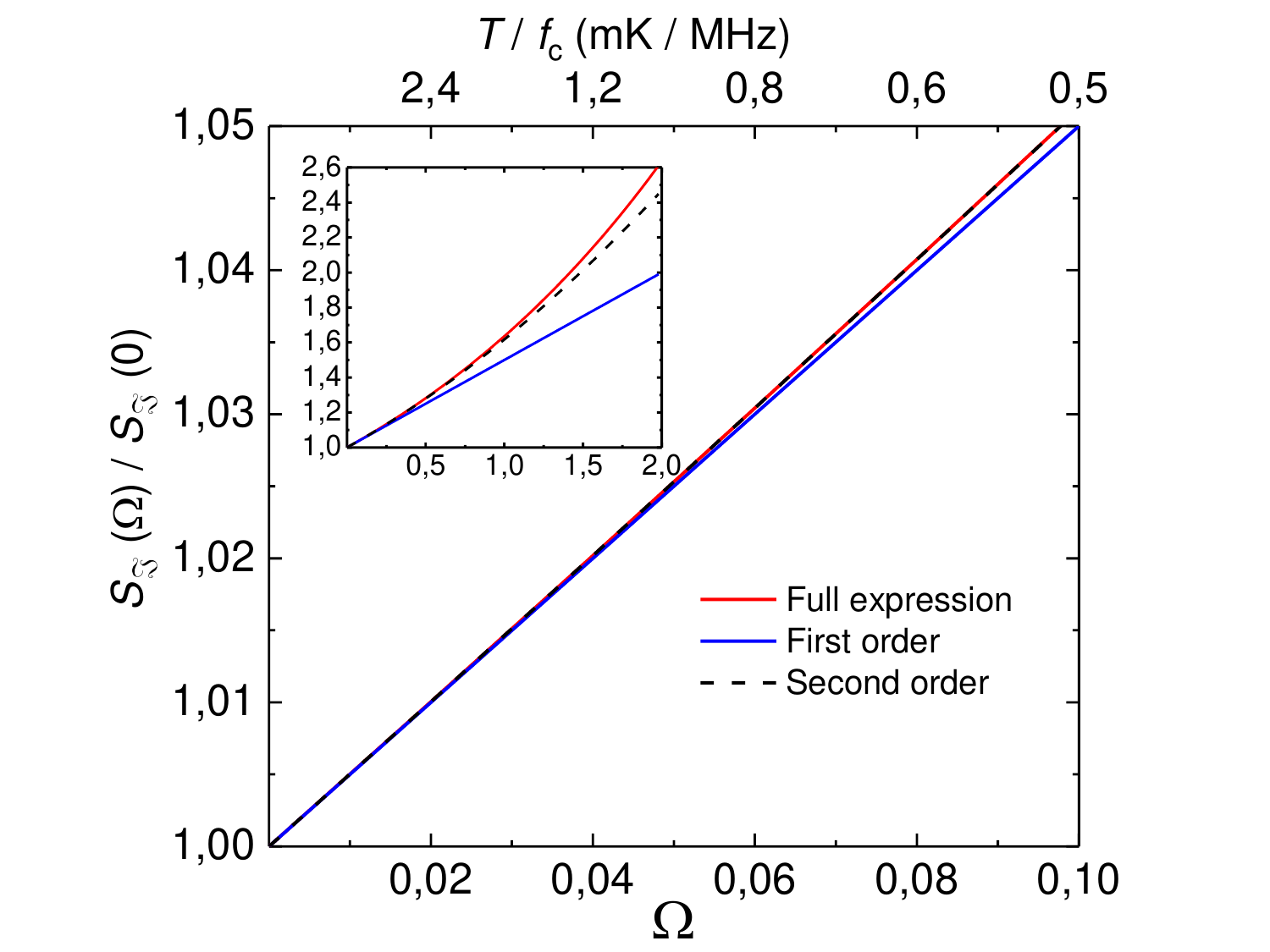}
\caption{Heat current noise $S_{\mathfrak{I}}(\Omega)$ normalized by its classical value $S_{\mathfrak{I}}(0)$, where $\Omega=\beta\hbar\omega$. Solid red line is the full expression based on Eq. (\ref{dimensionlesssIw2}); solid blue line shows the first order approximation $1+\frac{\Omega}{2}$ while dashed black line is the second order result based on Eq. (\ref{Sexpansion}). Top scale indicates the temperature $T$ with a detector of cut-off frequency $f_c$ for achieving the frequency on the bottom axis. Inset: The same data as in the main frame over a wider frequency range.}
\label{fig:2}       
\end{figure}
For zero temperature $\beta_p,\beta_e\rightarrow \infty$, we see that only the first term in the integrand of Eq. (\ref{sIw2}) is non-zero and just over the interval $0< \epsilon < \hbar \omega$.  Then, 
\begin{eqnarray} \label{nonzeroT}
S_{\mathfrak I}(\omega)= \frac{\Sigma \mathcal V}{96\zeta (5)k_B^5}\frac{(\hbar\omega)^6}{60}.
\end{eqnarray}
This result is in analogy with the expressions of zero-temperature noise of heat current in charge transport through a scatterer, for instance a tunnel junction \cite{pekolaPRL,sergi,hanggi}. In that case $S_{\mathfrak{I}}\propto \omega^3$, which is consistent with our result in the following way. The thermal conductance in a tunnel junction scales as $\propto T$. The fluctuation dissipation theorem then yields $S_{\mathfrak{I}}(0)\propto T^3$, to be compared to $T^6$ in the electron-phonon system (Eq. (\ref{nonzeroT})). Now, replacing $k_{\rm B}T$ by $\hbar\omega$ in the two cases, we get the corresponding zero temperature finite frequency noise $\propto \omega^3$ and $\propto \omega^6$, respectively.
\section{Discussion of fluctuations at finite frequency}
At equal temperatures and finite frequency, Eq. (\ref{sIw2}) can be written in the dimensionless form 
\begin{eqnarray} \label{dimensionlesssIw2}
S_{\mathfrak I}(\Omega)=\frac{S_{\mathfrak I}(0)}{960 \zeta (5)} \int_0^\infty du \,u^2 &&\Big[(2u-\Omega)^2\frac{1}{1-e^{-u}}\,\,\frac{u-\Omega}{e^{u-\Omega}-1}\nonumber\\&&+(2u+\Omega)^2\frac{1}{e^{u}-1}\,\,\frac{u+\Omega}{1-e^{-(u+\Omega)}}\Big],
\end{eqnarray}
where $\Omega\equiv \beta \hbar \omega$. The expansion of $S_{\mathfrak I}(\Omega)$ in $\Omega$ up to second order yields
\begin{equation}\label{Sexpansion}
\frac{S_{\mathfrak I}(\Omega)}{S_{\mathfrak I}(0)}=1+\frac{\Omega}{2}+(\frac{1}{12}+\frac{7}{240}\frac{\zeta (3)}{\zeta (5)})\Omega^2.
\end{equation}
We present the result of Eq. (\ref{dimensionlesssIw2}) for the noise power in the case of equal temperatures as a function of $\Omega$ in Fig. \ref{fig:2}. The analytic approximations up to the first and second order in $\Omega$ are also shown, and we see that the second order result follows the exact result up to $\Omega \sim 1$. 

We finally comment on the observability of the finite frequency corrections. Experimental techniques to measure temperature focus traditionally into the low frequency regime. Yet RF-techniques developed for charge transport \cite{schoelkopf} and circuit QED (Quantum Electro-Dynamics) experiments \cite{wallraff} have recently been adapted to thermometry to address temporal evolution of temperature and noise in heat currents down to sub-$\mu$s time scales \cite{schmidt,simone,olli}. With these methods, noise up to $f_c \sim 1...10$ MHz frequencies becomes experimentally feasible. The experimentally available $\Omega$ range $(0<\Omega< \Omega_{max})$ in Fig. \ref{fig:2}, can be obtained by setting $\Omega_{max}=hf_c/k_{\rm B}T$, where $T$ is the temperature of the experiment. We indicate $T$ on the top axis of Fig. \ref{fig:2} scaled by cut-off frequency $f_c$ of the thermometer. As an example we see that for $f_c=10$ MHz, one needs to measure at $T=5$ mK to achieve $\Omega_{max}=0.1$. This electronic temperature is within the range of present day experiments on nanostructures \cite{Iftikhar,anna}. In this case a correction of $\sim 5\%$ in $S_{\mathfrak I}$ can be observed. However, at the time when the experimental observation of even the classical heat current noise remains elusive, the experiment on quantum heat current noise is still a challenge.

We acknowledge funding from the Academy of Finland under grants 272218 and 273827. 

%
%





\end{document}